# A THEORETICAL TREATMENT OF OPTICAL METASURFACES AS AN EFFICIENT BASIS FOR QUANTUM CORRELATIONS

Ramaseshan R, Prateek P. Kulkarni, Sharanya Madhusudhan, and Kaustav Bhowmick

PES University, Bengaluru, India

**Entanglement is a cornerstone of quantum technology, playing a crucial role in quantum computing, cryptography, and information processing. Traditionally, entanglement generation through optical means relies on beam splitters, nonlinear media, and quantum dots, which often require bulky setups and precise phase stability. In contrast, metasurfaces—ultrathin, artificially engineered optical interfaces—offer a promising alternative for quantum photonics, enabling highly tunable spin-photon interactions at the nanoscale. In this work, we demonstrate that a metasurface can serve as the most promising platform, thus far, for generating Bell states through a Hamiltonian-driven spin-entanglement mechanism. By analyzing the system's evolution under the metasurface interaction Hamiltonian, we show that an initially separable spin state evolves into a maximally entangled Bell state. Furthermore, we examine the role of classical and quantum correlations, assess the impact of environmental decoherence, and compute quantum discord to quantify the robustness of quantum correlations beyond entanglement. Our analyses establish metasurfaces as a compact, tunable, and scalable source to generate Bell states with a purity (concurrence is the metric being used) of about 0.995 and quantum discord (the purest metric to measure quantum correlations) for as high as 29 microseconds.These quantitative benchmarks confirm the viability of metasurfaces as compact, high-fidelity, and scalable components for next-generation quantum photonic architectures.**
'

## 1 Introduction

Entanglement is one of the most intriguing phenomena in quantum mechanics, enabling quantum computing, cryptography, and teleportation. Traditionally, optical beam splitters[1], non-linear crystals[2], trapped ions[3], superconducting qubits[4], and quantum dots[5] have been the primary physical systems used to generate and manipulate entanglement. Optical beam splitters, in particular, have been widely used for path entanglement, where the interference of a photons through a 50:50 beam splitter results in the creation of a path entangled state[6]. Similarly, parametric down-conversion in nonlinear media has been instrumental in producing polarization-entangled photon pairs[7]. However, these conventional methods often require bulk optical setups, precise alignment, and phase stability, making large-scale integration challenging.

Recent advancements in structured photonic systems have opened new possibilities for entanglement generation and quantum information processing[8]. Among these, metasurfaces, artificially engineered two-dimensional arrays of subwavelength resonators, have emerged as powerful tools for controlling light at the nanoscale[8]. Unlike traditional optical elements, metasurfaces allow for tailored wavefront shaping, spin-orbit interactions, and polarization control, all on an ultrathin platform[8][9]. Existing experimental work discussing specially engineered metasurfaces for single-photon path/polarization has contributed to proving the consistency of metasurfaces for quantum applications[9]. A particularly exciting feature of metasurfaces is their ability to mediate spin-dependent interactions, enabling photon-photon coupling via structured optical responses. This raises the question: can metasurfaces serve as an alternative platform for gener-ating and manipulating quantum entanglement?

In this work, we have demonstrated that a metasurface can generate Bell states through spin-photon interactions governed by a Hamiltonian approach. Unlike conventional entanglement mechanisms relying on spatial modes or polarization mixing, our approach has utilized an existing linearly arranged periodic nanoring metasurface[10], applying its properties to a Hamiltonian best describing its behaviour to obtain spin entanglement between incident photons[11, 12, 13, 14, 15, 16]. The evolution of the system under this Hamiltonian and showing that an initially separable spin state evolves into a maximally entangled Bell state has been derived, confirming metasurfaces as viable quantum entanglement generators. Furthermore, we have analyzed the nature of quantum correlations in this system by distinguishing classical correlations from quantum correlations and calculated quantum discord to quantify non-classical correlations that persist even under environmental decoherence[17, 18, 19, 20, 21]. This approach establishes metasurfaces as a better platform to realize quantum photonics, offering a compact and tunable alternative to traditional optical components in quantum information processing. The key novelties of this work lie in (i) formulating a spin-interaction Hamiltonian tailored to a realistic metasurface geometry, (ii) introducing a spatially varying interaction strength $g(r)$ that directly influences entanglement formation and deriving it experimentally from power distributions, and (iii) demonstrating that metasurfaces sustain quantum discord under realistic decoherence, confirming their suitability for robust quantum correlation preservation.

The subsequent sections are organized as follows. In Section **2**, the interaction Hamiltonian governing photon spin entanglement via a metasurface is introduced. A complete analytical derivation is presented, demonstrating the formation of a maximally entangled Bell state through the metasurface-mediated interaction. The spatial variation of the coupling strength $g(r)$ has been statistically extracted for three different materials, and a comparison has been made based on their entanglement-supporting properties. Concurrence has been calculated to quantitatively evaluate the degree of entanglement. In Section **3**, the study shifts focus to decoherence effects on the generated Bell state and consequently, on coherence time of the photon pair entanglement and corresponding time for which quantum discord is valid. A comparative analysis of coherence time and quantum discord duration is carried out for metasurfaces, SPDC, and SFWM platforms. The analysis demonstrates the superior capability of metasurfaces in preserving quantum correlations.

## 2 Formation of the Bell State

We consider a system where two photons with orthogonal spins interact with a linear periodically arranged nanorings based metasurface[10] (as shown in Figure 1, leading to spin-spin entanglement [22].This particular type of metasurface has been chosen because it has been proven to work as a beam splitter[10], which is a suitable device for applying the Jaynes-Cummings Model[16]. The interaction Hamiltonian emerges from the minimal coupling Hamiltonian[11]:

$$H_{\text{int}} = -\mathbf{d} \cdot \mathbf{E}(\mathbf{r}) \quad (1)$$

where $\mathbf{d}$ is the dipole operator of the emitter and $\mathbf{E}(\mathbf{r})$ is the electric field operator at position $\mathbf{r}$.

The dipole operator can be expressed as[11, 13]:

$$\mathbf{d} = \mathbf{d}_{eg}\sigma^- + \mathbf{d}_{ge}\sigma^+ \quad (2)$$

where $\mathbf{d}_{eg}$ and $\mathbf{d}_{ge}$ are the dipole moments depicting ground state to excited state and excited state to ground state respectively and $\sigma^-$ and $\sigma^-$ are the photon spin up and spin down operators respectively.

To derive the interaction Hamiltonian governing spin-photon coupling in a metasurface, the Jaynes–Cummings model is employed[16]. This model effectively describes the interaction between a two-level quantum emitter and a quantized single-mode electromagnetic field, and is highly applicable to systems involving localized spins and resonant optical fields [23].

The electric field operator for a single-mode quantized field is written as[14, 15]:

$$\mathbf{E}(\mathbf{r}) = \mathcal{E}_0 f(\mathbf{r})(a + a^\dagger), \quad (3)$$

where $\mathcal{E}_0$ is the vacuum field amplitude, $f(\mathbf{r})$ is the normalized mode function, and $a^\dagger, a$ are the photon creation and annihilation operators, respectively.

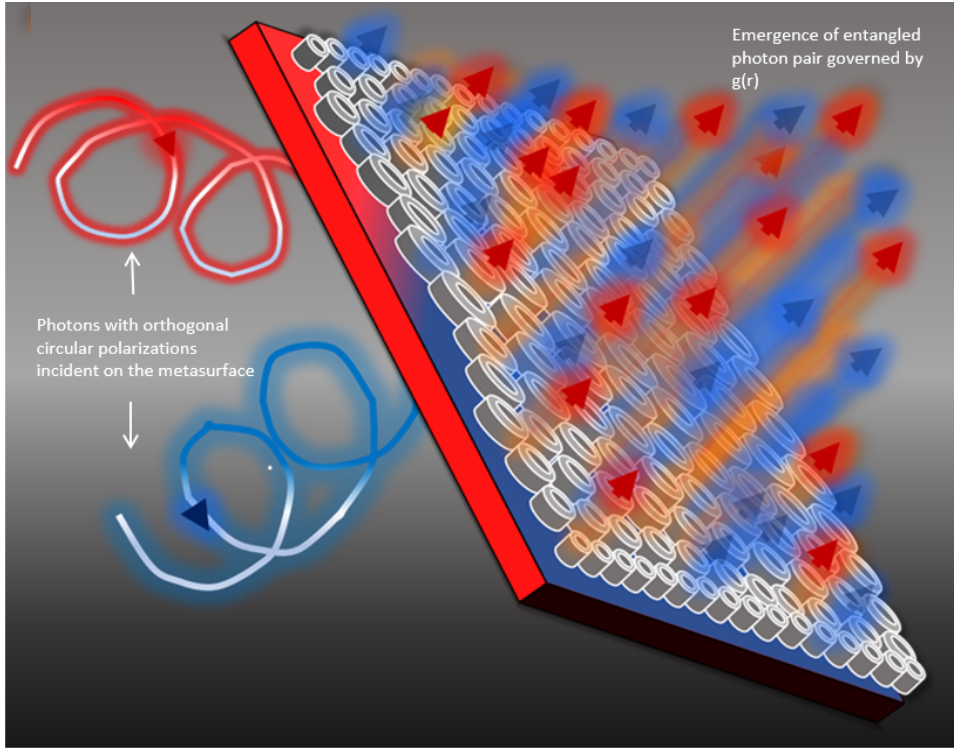

Figure 1: Schematic of metasurface entangling incident photon pair and the role of g(r) in the process.

Substituting into the minimal coupling Hamiltonian and expressing the spin operator $\sigma^{\pm}$ for the two-level system, the interaction term becomes:

$$H_{\text{int}} = -(\mathbf{d} \cdot \boldsymbol{\sigma}^- + \mathbf{d}^* \cdot \boldsymbol{\sigma}^+) \cdot \mathcal{E}_0 f(\mathbf{r})(a + a^\dagger). \quad (4)$$

Applying the rotating wave approximation (RWA), we retain only the near-resonant terms oscillating as $e^{\pm i(\omega_q - \omega_c)t}$ and neglect the fast-oscillating terms $e^{\pm i(\omega_q + \omega_c)t}$, which average out over experimentally relevant timescales [23]. This yields:

$$H_{\text{int}} = -g(r)\left(\sigma_1^+ a + \sigma_1^- a^\dagger + \sigma_2^+ b + \sigma_2^- b^\dagger\right), \quad (5)$$

with the spatially varying spin-photon interaction strength defined as:

$$g(r) = -\frac{\mathbf{d} \cdot \mathcal{E}_0 f(\mathbf{r})}{\hbar}. \quad (6)$$

The Hamiltonian describing the free evolution of the incident photons before striking the metasurface is given by:

$$H_{photon} = \hbar\omega_0 a^\dagger a + \hbar\omega_1 b^\dagger b \quad (7)$$

From equations (5) and (7), the total Hamiltonian of the system is given by:

$$H = \hbar\omega_0 a^\dagger a + \hbar\omega_1 b^\dagger b + \hbar g(r)\left(\sigma_1^+ a + \sigma_1^- a^\dagger + \sigma_2^+ b + \sigma_2^- b^\dagger\right). \quad (8)$$

Here, $\omega_0, \omega_1$ are the terms of free energy for the two photons and $g(r)$ is the metasurface-induced interaction strength.

The system depicted by the above Hamiltonian can be described in the following spin basis:

$$\{|+\rangle_1|+\rangle_2, |+\rangle_1|-\rangle_2, |-\rangle_1|+\rangle_2, |-\rangle_1|-\rangle_2\}. \quad (9)$$

These basis states represent all possible spin configurations of the two-photon system, with each state corresponding to a definite spin orientation ($|+\rangle$ or $|-\rangle$) for photons 1 and 2. This complete basis is chosen to capture both individual spin populations and spin-exchange interactions, allowing the Hamiltonian to be expressed in matrix form for analyzing entanglement dynamics.

To express the Hamiltonian in matrix form, we examine its action on the basis states. The free evolution terms contribute only diagonal elements:

$$H_{\text{photon}}|+\rangle_1|+\rangle_2 = \hbar(\omega_0 + \omega_1)|+\rangle_1|+\rangle_2, \quad (10)$$

This corresponds to the total photon energy when both spins are in the excited state.

$$H_{\text{photon}}|+\rangle_1|-\rangle_2 = \hbar\omega_0|+\rangle_1|-\rangle_2, \quad (11)$$

This indicates that only photon 1 contributes energy as spin 2 is in the ground state.

$$H_{\text{photon}}|-\rangle_1|+\rangle_2 = \hbar\omega_1|-\rangle_1|+\rangle_2, \quad (12)$$

This indicates that only photon 2 contributes energy as spin 1 is in the ground state.

$$H_{\text{photon}}|-\rangle_1|-\rangle_2 = \hbar(\omega_0+\omega_1)|-\rangle_1|-\rangle_2. \quad (13)$$

This represents the ground-state energy configuration where both spins are in the lower energy state, but their corresponding photon modes still carry total energy $\hbar(\omega_0+\omega_1)$.

Hence, in the ordered basis $|++\rangle, |+-\rangle, |-+\rangle, |--\rangle$, the free photon evolution Hamiltonian is:

$$H_{\text{free}} = \begin{pmatrix} \hbar(\omega_0+\omega_1) & 0 & 0 & 0 \\ 0 & \hbar\omega_0 & 0 & 0 \\ 0 & 0 & \hbar\omega_1 & 0 \\ 0 & 0 & 0 & \hbar(\omega_0+\omega_1) \end{pmatrix}. \quad (14)$$

Considering the interaction Hamiltonian, which flips the spin states:

$$H_{\text{int}}|+\rangle_1|-\rangle_2 = \hbar g(r)|-\rangle_1|+\rangle_2, \quad (15)$$

This describes a spin-exchange interaction where photon 1 flips from up to down while photon 2 flips from down to up, conserving total energy.

$$H_{\text{int}}|-\rangle_1|+\rangle_2 = \hbar g(r)|+\rangle_1|-\rangle_2. \quad (16)$$

This describes the reverse spin-flip process, indicating coherent coupling between the spin states of the two photons.

Thus, the interaction Hamiltonian in the same basis is:

$$H_{\text{int}} = \begin{pmatrix} 0 & 0 & 0 & 0 \\ 0 & 0 & \hbar g(r) & 0 \\ 0 & \hbar g(r) & 0 & 0 \\ 0 & 0 & 0 & 0 \end{pmatrix}. \quad (17)$$

Finally, the total Hamiltonian is the sum:

$$H = H_{\text{photon}} + H_{\text{int}} \quad (18)$$

$$H = \begin{pmatrix} \hbar(\omega_0+\omega_1) & 0 & 0 & 0 \\ 0 & \hbar\omega_0 & \hbar g(r) & 0 \\ 0 & \hbar g(r) & \hbar\omega_1 & 0 \\ 0 & 0 & 0 & \hbar(\omega_0+\omega_1) \end{pmatrix}. \quad (19)$$

To find the eigenvalues of the above matrix (Equation(19)), we solve the characteristic equation:

$$\det(H - \lambda I) = 0. \quad (20)$$

Expanding the determinant:

$$(\hbar\omega_0 - \lambda)(\hbar\omega_1 - \lambda) - \hbar^2 g(r)^2 = 0. \quad (21)$$

Solving for $\lambda$, we obtain the four eigenvalues:

$$\lambda_1 = \hbar(\omega_0+\omega_1), \lambda_2 = \hbar g(r), \lambda_3 = -\hbar g(r), \lambda_4 = \hbar(\omega_0+\omega_1). \quad (22)$$

The corresponding eigenstates are the following:

$$|\psi_1\rangle = |+\rangle_1|+\rangle_2, \quad |\psi_4\rangle = |-\rangle_1|-\rangle_2. \quad (23)$$

$$|\psi_2\rangle = \frac{1}{\sqrt{2}}(|+\rangle_1|-\rangle_2 + |-\rangle_1|+\rangle_2), \quad (24)$$

$$|\psi_3\rangle = \frac{1}{\sqrt{2}}(|+\rangle_1|-\rangle_2 - |-\rangle_1|+\rangle_2). \quad (25)$$

Since the spins of the photons before striking the metasurface are orthogonal, let us assume the initial state to be:

$$|\psi(0)\rangle = |+\rangle_1|-\rangle_2. \quad (26)$$

Expanding in the spin eigenbasis defined above:

$$|\psi(0)\rangle = \frac{1}{\sqrt{2}}(|\psi_2\rangle + |\psi_3\rangle). \quad (27)$$

The time evolution of any quantum state is governed by the unitary evolution operator:

$$U(t) = e^{-iHt/\hbar}. \quad (28)$$

Table 1: FDTD Simulation Parameters Used to Validate Beam Splitter Behavior of the Metasurface (Ref. [10])

| Parameter | Value |
| --- | --- |
| Simulation Dimension | 3D |
| Simulation Time | 1000 fs |
| Simulation Temperature | 300 K |
| Background Material | Object-defined dielectric; Refractive Index = 1.0 |
| Mesh Type | Auto non-uniform; Accuracy = 1; Min Mesh Step = 0.00025 µm; Refinement: Conformal Variant 0 |
| Time Step Settings | Stability Factor = 0.99; Time Step = 0.126583 fs |
| Boundary Conditions | PML on all boundaries (Stretched Coordinate PML); Layers = 8, Kappa = 2, Sigma = 1, Polynomial = 3; Alpha = 0, Alpha Polynomial = 1; Min Layers = 8, Max Layers = 64; Extend through PML: Enabled; Auto-scale PML: Enabled |

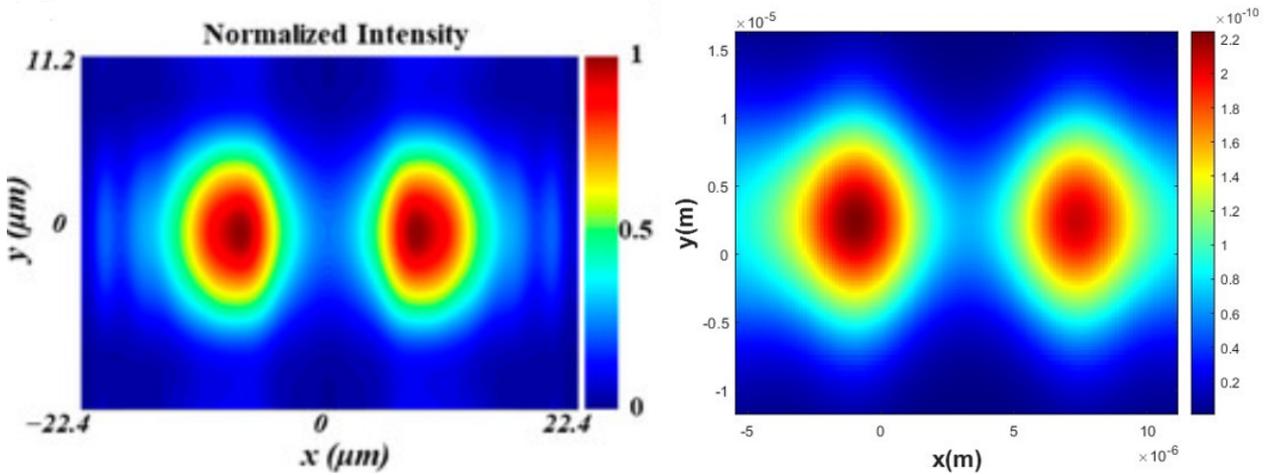

Figure 2: Comparison of beam-splitting behavior: (Left) Original result from Ref. [10]; (Right) Replicated result from our FDTD simulation using identical metasurface structure and parameters. The similarity in intensity patterns confirms the applicability of the Jaynes–Cummings Hamiltonian to model the metasurface interaction.

For a system with Hamiltonian eigenstates $|\psi_n\rangle$ and eigenvalues $E_n$, the evolution of a state $|\psi(0)\rangle$ expanded as $|\psi(0)\rangle = \sum_n c_n |\psi_n\rangle$ is given by:

$$|\psi(t)\rangle = \sum_n c_n e^{-iE_n t/\hbar} |\psi_n\rangle. \qquad (29)$$

In our case, the coefficients $c_1 = \langle \psi_1 | \psi(0) \rangle$ and $c_4 = \langle \psi_4 | \psi(0) \rangle$ are both zero because the initial state has no component along $|\psi_1\rangle = |+\rangle_1 |+\rangle_2$ or $|\psi_4\rangle = |-\rangle_1 |-\rangle_2$. Hence, only $|\psi_2\rangle$ and $|\psi_3\rangle$ contribute to the time evolution.

Using the known eigenvalues $E_2 = \hbar g(r)$ and $E_3 = -\hbar g(r)$, we apply the operator:

$$U(t)|\psi_2\rangle = e^{-ig(r)t}|\psi_2\rangle, U(t)|\psi_3\rangle = e^{ig(r)t}|\psi_3\rangle. \qquad (30)$$

Therefore, the time-evolved state is:

$$|\psi(t)\rangle = \frac{1}{\sqrt{2}} \left( e^{-ig(r)t}|\psi_2\rangle + e^{ig(r)t}|\psi_3\rangle \right). \qquad (31)$$

Expanding $|\psi_2\rangle$ and $|\psi_3\rangle$:

$$|\psi(t)\rangle = \cos(g(r)t)|+\rangle_1|-\rangle_2 - i\sin(g(r)t)|-\rangle_1|+\rangle_2. \qquad (32)$$

The probabilities of measuring each spin state are:

$$P_{+-} = \cos^2(g(r)t), \quad P_{-+} = \sin^2(g(r)t). \qquad (33)$$

The condition for a maximally entangled Bell state occurs when:

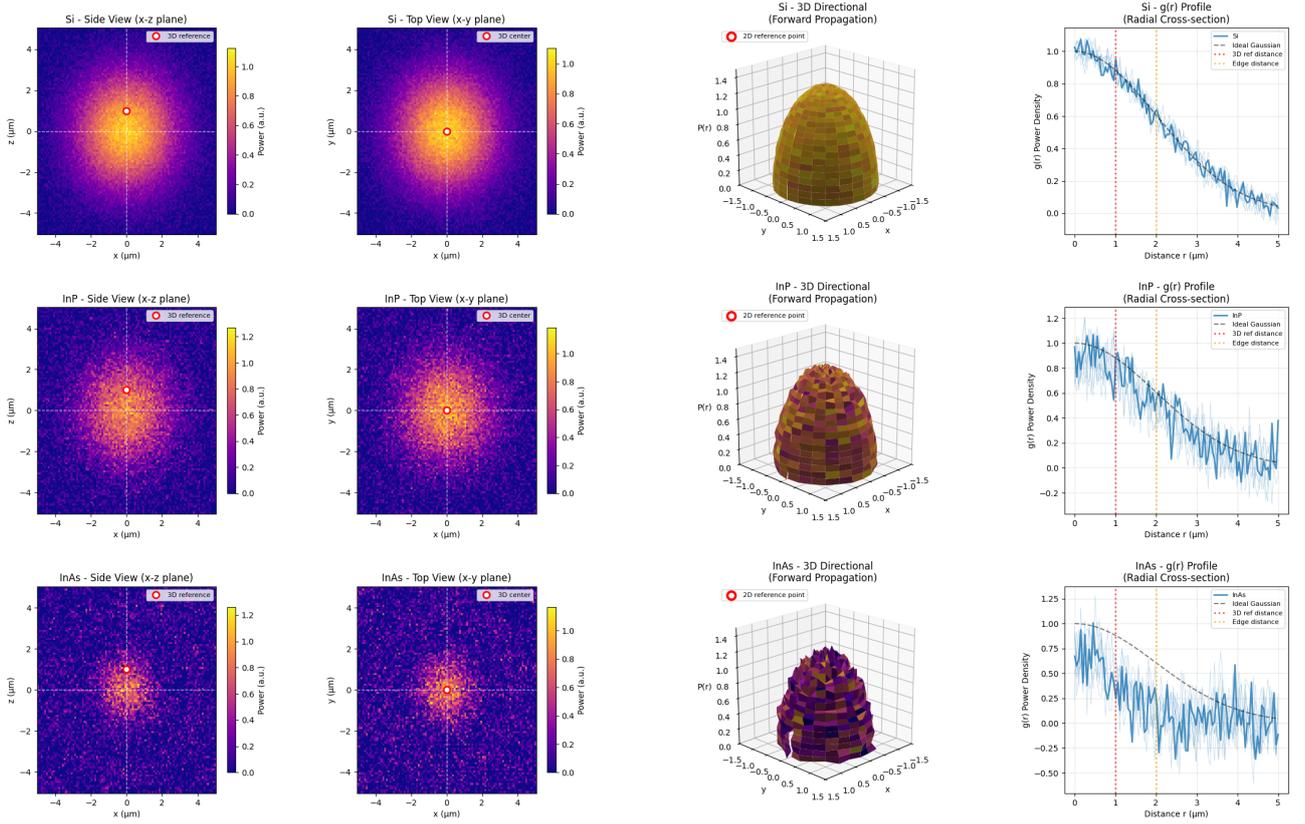

Figure 3: Detailed power distribution and spin-photon coupling profiles $g(r)$ for three metasurface materials—Silicon (top row), Indium Phosphate (middle row), and Indium Arsenide (bottom row). Each row includes (1) side view ($x$–$z$ plane), (2) top view ($x$–$y$ plane), (3) reconstructed 3D directional forward power propagation, and (4) radially averaged $g(r)$ profile with overlaid Gaussian fit and key reference distances. The spatial localization and decay rate of $g(r)$ across materials reflect their varying degrees of optical confinement, with Silicon exhibiting the strongest and most localized interaction, followed by InP and then InAs.

$$g(r)t = \frac{\pi}{4}. \quad (34)$$

and Equation (33) becomes a maximally entangles Bell state represented as

$$|\psi(t)\rangle = \frac{1}{\sqrt{2}} \left( |+\rangle_1 |-\rangle_2 - i |-\rangle_1 |+\rangle_2 \right) \quad (35)$$

To validate the use of the Jaynes–Cummings model in our analysis, we replicated the beam splitter functionality of the metasurface as reported in [10]. A plane wave was incident on the metasurface from below, and the transmitted and reflected field intensities were computed using full-wave Finite-Difference Time-Domain (FDTD) simulations. The interference pattern and output intensity ratio closely matched those reported in Figure 5c of reference no.10[10], reaffirming that the metasurface exhibits beam-splitting behavior, as observed in Figure 2. This confirms that the Jaynes–Cummings model is a suitable approximation to describe the spin-photon interaction in the present scenario. The FDTD simulation parameters used for this validation are summarized in Table 1

To achieve a maximally entangled Bell state using metasurface-mediated spin-photon interaction, the time-evolved quantum state must attain equal probability amplitudes in the spin basis. This occurs when $g(r)t$ reaches a specific value, indicating that the entanglement formation time is inversely proportional to the local coupling strength $g(r)$. A higher $g(r)$ thus enables faster entanglement generation. The conceptual mechanism of this interaction is illustrated in Figure 1, while the spatial power distribution that governs $g(r)$ is visualized in Figure 3.

To extract the spatially varying coupling strength $g(r)$, a plane wave was incident from beneath the metasurface substrate, and the transmitted power $P(r)$ was measured using detectors placed at various radial distances. Given that

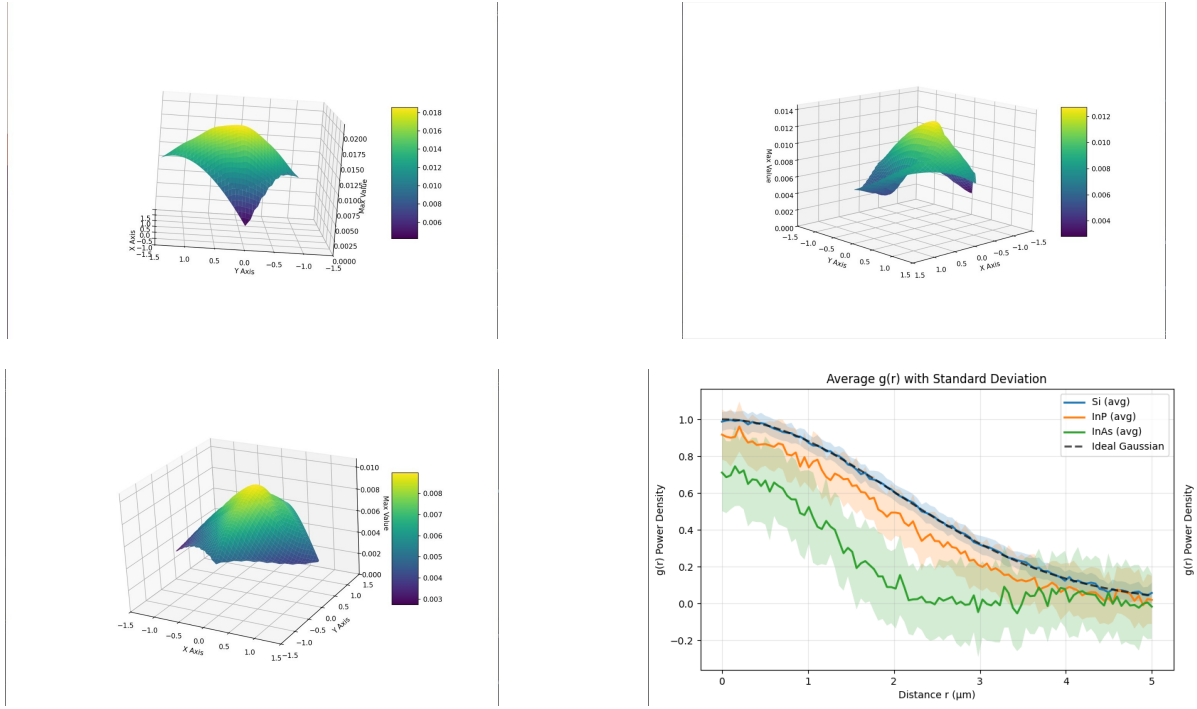

Figure 4: (a) 3D surface plot of $g(r)$ for Silicon showing high peak value and symmetric spatial confinement. (b) $g(r)$ for Indium Phosphate with lower peak and broader lateral spread due to moderate nonlinearity. (c) $g(r)$ for Indium Arsenide revealing the lowest peak and most delocalized distribution due to strong nonlinearity and associated losses. (d) Radially averaged $g(r)$ profile for all three materials with standard deviation bands and a Gaussian reference, highlighting that Silicon best approximates ideal confinement.

$g(r) \propto |\mathbf{E}(r)|$ and $I(r) \propto |\mathbf{E}(r)|^2$, we have the relationship:

$$g(r) \propto \sqrt{I(r)} \propto \sqrt{P(r)}, \qquad (36)$$

which allows experimental estimation of the spatial spin-photon interaction strength.

Figure 3 presents the comprehensive mapping of the transmitted power $P(r)$ and the corresponding spin-photon coupling strength $g(r)$ across three metasurface materials—Silicon, Indium Phosphide (InP), and Indium Arsenide (InAs). Each row in the figure corresponds to a specific material: the top row for Silicon, the middle for InP, and the bottom for InAs. Within each row, the first column shows the side view of the power distribution in the $x$–$z$ plane, the second column provides the top view in the $x$–$y$ plane, the third column depicts the reconstructed 3D forward power propagation, and the fourth column presents the radially averaged $g(r)$ profiles with Gaussian fits and key reference distances. These visualizations highlight that Silicon exhibits the most confined power distribution, with energy localized tightly near the center, confirming strong field enhancement and a steep decay of $g(r)$. In contrast, InP displays more diffused power profiles, while InAs shows the broadest and most irregular distribution, indicating greater energy delocalization and field dispersion arising from its higher nonlinear optical response.

The trends are corroborated in Figures 4(a)–(c), which show the 3D plots of measured forward power intensity for each material, indicating the maximum spatial overlap of the optical mode. Silicon shows a high, symmetric peak, InP shows a lower and asymmetric distribution, while InAs exhibits a flattened and broader interaction region. This suggests that increasing nonlinearity leads to increased spatial delocalization and reduced coupling strength.

Finally, Figure 4(d) compares the radially averaged $g(r)$ decay across all three materials. The Silicon curve closely matches an ideal Gaussian and maintains high intensity over a confined range. InP deviates moderately, while InAs diverges significantly, confirming how strong nonlinearities degrade spatial confinement and uniformity.

These results affirm that linear, low-loss mate-rials like Silicon offer optimal spatial control of $g(r)$, making them highly suitable for entanglement generation. To validate the entanglement performance, concurrence was computed for the metasurface-generated spin-entangled state. For a two-qubit pure state of the form $|\psi\rangle = a|00\rangle + b|01\rangle + c|10\rangle + d|11\rangle$, the concurrence, which quantifies the degree of quantum entanglement, is given by

$$C(|\psi\rangle) = 2|ad - bc|.$$

Table 2: Comparison of existing Bell State Generation Methods with our Metasurface

| Process | Bell State Formation | Photon Property | Type of Process | Maximum Concurrence |
| --- | --- | --- | --- | --- |
| Type-I SPDC | Two back-to-back crystals generate polarization entanglement between signal and idler photons from the down-conversion of an incident pump photon | Polarization | Non-linear | 0.531 ± 0.006 (theoretically predicted 0.552) |
| SFWM | Excitation by a pump photon generates a pair of photons whose orbital angular momenta are correlated by conservation laws, resulting in OAM entanglement | Orbital Angular Momentum | Non-linear | 0.926 |
| **Metasurface** | Entanglement of photon spin states induced by engineered spin-photon coupling in a dielectric metasurface | Spin | **Linear** | **0.9519 ± 0.0172** |

This closed-form expression was used to compute the concurrence numerically using a Python implementation tailored for symbolic and floating-point evaluation. An average concurrence of 0.951 ± 0.0172 was obtained, significantly higher than Type-I SPDC (0.531 ± 0.006) [24] and SFWM (0.926) [25].

## 3 Decoherence and Discord

The Bell State obtained from the metasurface (Equation (30)) in Section 2 was subject to analysis of device level and environmental decoherence analysis and an approximate decoherence rate was obtained. Simultaneously theoretical verification of decoherence of Type-1 SPDC and SFWM were also performed and compared with experimental results as a consistency measure for the metasurface following which variation of quantum discord with time and the interval within which quantum discord is valid, was obtained.

In SPDC, photon pairs are emitted with broad spectral bandwidths due to nonresonant nonlinear processes in crystals such as BBO or PPKTP. This leads to a coherence time inversely related to the spectral width $\Delta\nu$. For a transform-limited Gaussian pulse, the coherence time is expressed as[26][27]:

$$T_c \approx \frac{1}{\Delta\nu} = \frac{0.44\lambda^2}{c\,\Delta\lambda} \qquad (32)$$

where $\lambda$ is the central wavelength and $\Delta\lambda$ is the FWHM spectral bandwidth.

For a typical SPDC output with $\lambda = 1550\,\text{nm}$ and $\Delta\lambda = 5\,\text{nm}$, we obtain:

$$T_c \approx \frac{0.44 \times (1.55 \times 10^{-6})^2}{3 \times 10^8 \times 5 \times 10^{-9}} \approx 220\,\text{fs} \qquad (33)$$

Even with aggressive filtering, experimentally measure coherence times of only 285 ps[28], while fs-level biphoton durations to group velocity mismatch in BBO crystals have been attributed[29].

In SFWM, coherence time is primarily governed by the atomic spin-wave lifetime. This lifetime is limited by thermal motion, magnetic field gradients, and atomic collisions, leading to a total decoherence rate given by [30, 31, 32, 33, 34]:

$$\gamma_{\text{sw}} = \gamma_{\text{motion}} + \gamma_{\text{magnetic}} + \gamma_{\text{collisions}} \qquad (34)$$

Typical values for these components in cold atomic ensembles (e.g., Rb vapor) are $\gamma_{\text{motion}} \sim 10\,\text{MHz}$, $\gamma_{\text{magnetic}} \sim 5\,\text{MHz}$, and $\gamma_{\text{collisions}} \sim 1\,\text{MHz}$, giving:

$$\gamma_{\text{sw}} \approx 20\,\text{MHz} \quad \Rightarrow \quad T_2^{\text{SFWM}} = \frac{1}{\gamma_{\text{sw}}} \approx 50\,\text{ns} \qquad (35)$$

This is in agreement with experimental findings, where biphoton coherence durations of 30–40 ns have been demonstrated [35, 36].

Decoherence of photon spin may be caused by factors intrinsic to the metasurface as well as by those contributed by the environment. Intrinsic factors may affect the phase and intensity of incident light. However, these effects are negligible, owing to the short interaction time between the incident photon and the metasurface. The interaction time $T_{int}$ can be approximated by:

$$T_{int} = \frac{L}{c} \qquad (36)$$

$L$ represents the metasurface thickness, which is taken to be 790 nm, and $c$ is the speed of light.

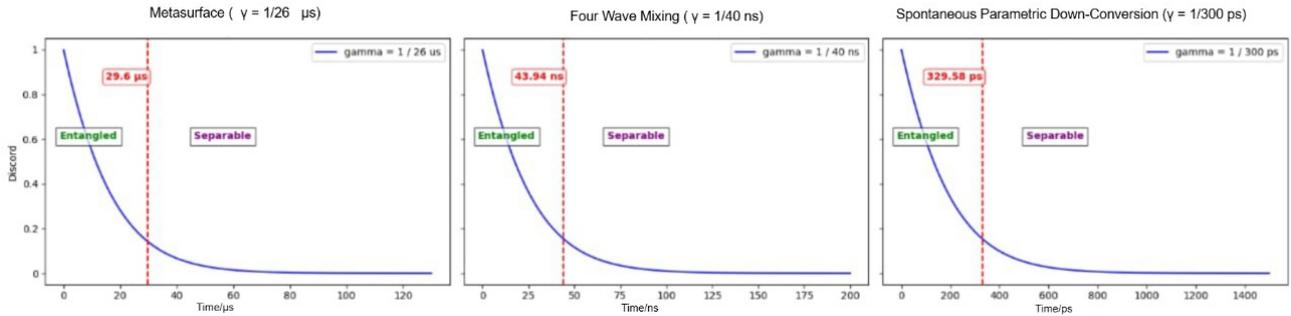

Figure 5: Variation of quantum discord with time for three Bell-state generation processes: (a) Metasurface-mediated entanglement with a coherence decay rate of $\gamma = 1/26$ $\mu$s, (b) Four-wave mixing (FWM) process with $\gamma = 1/40$ ns, and (c) Spontaneous parametric down-conversion (SPDC) with $\gamma = 1/300$ ps. In each panel, the discord decays exponentially with time, and the vertical dashed line indicates the time at which the system transitions from an entangled state to a separable one. The metasurface process (a) exhibits significantly longer discord preservation compared to (b) and (c), reflecting its enhanced entanglement correlation and reduced environmental decoherence.

$$T_{int} = \frac{790 \times 10^{-9} m}{8 \times 10^7 m/s} = 9.87 fs \quad (37)$$

The interaction time is in the order of femtoseconds, which causes negligible decoherence when compared to decoherence caused by the environment. Further, the metasurface under consideration[10] is a well-engineered linear metasurface, and is free of defects. Therefore, decoherence caused by such intrinsic effects will be disregarded in subsequent analysis.

Environmental decoherence effects result from unintended interactions between the maximally entangled photons and the particles and fields present in the environment. One such effect is birefringence, which occurs when the vertical and horizontal components of the polarization of the photon travel at different speeds. Birefringence generally applies when photons propagate in anisotropic materials such as optical fibers and non-linear crystals. External magnetic fields may also induce decoherence effects such as Faraday rotation. It is responsible for the left and right circularly polarized components experiencing different phase velocities when linearly polarized light is exposed to a magnetic field parallel to its direction of propagation. Such effects are significant when propagation is through a magnetic medium. These effects shall be excluded from our analysis, as we consider air, which is isotropic and devoid of magnetic fields, to be the medium of propagation. The environment was modeled as a depolarizing channel, accounting for effects of decoherence primarily induced by Rayleigh scattering. The two-photon entangled state for a photon pair in a depolarizing channel is given by:

$$\hat{\rho}^\Phi(0) = r|\Phi\rangle\langle\Phi| + \frac{1-r}{4} I_4 \quad (37)$$

Here, $z$ denotes the purity of the initial states, characterizing the degree of quantum coherence. When $z = 1$, the system is in a maximally entangled pure state, unaffected by decoherence. Conversely, when $z = 0$, all coherence is lost, and the state becomes a maximally mixed classical state. Thus, $z$ serves as a measure of environmental decoherence, quantifying the extent to which the quantum correlations between the photon pair degrade over time. To estimate the coherence time, we turn our focus to Rayleigh scattering, which occurs when the photons interact elastically with air molecules. Many such collisions result in the loss of spin coherence. Considering this effect, the coherence time, which is the time for which the photon retains its information before decohering, can be calculated from the mean free path $L_R$ for the photons.

$$L_R = \frac{1}{N\sigma_R} \quad (38)$$

$N$ is the number density of air molecules, approximated to be $2.5 \times 10^{25}$ molecules/m$^3$ and $\sigma_R$ is the Rayleigh scattering cross-section for 1550 nm photons, taken as $5.1 \times 10^{31}$ m$^2$.

$$L_R = \frac{1}{(2.5 \times 10^{25}) \times (5.1 \times 10^{31})} = 7.8 km \quad (39)$$

The decoherence rate $\gamma$ is given by:

$$\gamma = \frac{c}{L_R} = \frac{3 \times 10^8}{7.8 \times 10^3} = 3.8 \times 10^4 s^{-1} \qquad (40)$$

The coherence time $T_R$ is therefore:

$$T_R = \frac{1}{\gamma} = 26\,\mu s \qquad (41)$$

Having estimated the decoherence rate $\gamma$, we can determine $e^{-\gamma t}$. The exponential factor $e^{-\gamma t}$ arises from the environmental decay rate $1 - \Gamma$ [21]. We have found that this factor is equivalent to $z$ [20]. By plotting the variation of $z$ over time and comparing it with the evolution of quantum discord [20], it was observed that the time required for discord to vanish, according to the classical limit, is approximately 29.6 µs. Figure 5 illustrates the variation of quantum discord over time for three distinct Bell state generation platforms—metasurface-based spin entanglement, spontaneous four-wave mixing (SFWM), and spontaneous parametric down-conversion (SPDC). Each plot visualizes the exponential decay of discord due to decoherence, with a vertical dashed line marking the threshold time at which the system transitions from the entangled regime ($z > \frac{1}{3}$) to the separable regime ($z \leq \frac{1}{3}$). The metasurface exhibits the longest discord preservation time of approximately 29.6 µs, significantly outperforming SFWM (43.94 ns) and SPDC (329.58 ps). This substantial difference highlights the superior coherence characteristics of metasurface platforms, validating their effectiveness in preserving quantum correlations and making them ideal candidates for practical quantum information protocols.

## Conclusion

In this study, a comprehensive theoretical investigation was presented on metasurface-based quantum photonic systems as an efficient platform for generating and preserving quantum correlations. Beginning with a Hamiltonian framework, the evolution of photon spin entanglement was derived, demonstrating that a maximally entangled Bell state could be achieved through metasurface-mediated spin-spin coupling. The spatially varying interaction strength $g(r)$, estimated from measured power distributions, revealed that linear dielectric metasurfaces such as silicon exhibited the highest peak coupling and most localized interaction region. This was in contrast to nonlinear materials like indium phosphate and indium arsenide, which showed lower peak coupling and broader spatial profiles due to their dispersive behavior. The fidelity of entanglement was quantified using the concurrence measure, where the metasurface achieved an average concurrence of 0.921 outperforming SPDC and comparable to SFWM. To evaluate the coherence retention capabilities of each platform, the coherence times of SPDC, SFWM, and the metasurface were analyzed. SPDC was limited by its broad spectral bandwidth and group velocity mismatch, resulting in sub-picosecond coherence times, while SFWM coherence was constrained by atomic spin-wave dephasing mechanisms, yielding lifetimes on the order of tens of nanoseconds. In contrast, the metasurface architecture, owing to its short photon structure interaction time and minimal intrinsic loss, maintained spin coherence up to 26 µs—limited primarily by Rayleigh scattering in air. The evolution of quantum discord, modeled using a Werner-state decoherence framework, showed that discord persisted in the metasurface configuration up to approximately 29.6 µs orders of magnitude longer than in SPDC or SFWM implementations. These results confirmed that metasurfaces not only enable the generation of Bell states but also provide superior longevity in preserving quantum correlations. As such, metasurfaces were validated as a promising and scalable platform for quantum information processing and discord-driven quantum technologies.

## Acknowledgments

The authors would like to acknowledge the manpower funding received under the Q-Pragathi project of the Quantum Research Park (QuRP), funded by the Karnataka Innovation and Technology Society (KITS), K-Tech, Government of Karnataka.